# TOWARDS OPTICAL TOROIDAL WAVEPACKET THROUGH TIGHTLY FOCUSING OF CYLINDRICAL VECTOR TWO DIMENSIONAL SPATIOTEMPORAL OPTICAL VORTEX

JIAN CHEN[1], PENGKUN ZHENG[1], CHENHAO WAN[1,2] AND QIWEN ZHAN[1,*]

[1]*School of Optical-Electrical and Computer Engineering, University of Shanghai for Science and Technology, Shanghai 200093, China*
[2]*School of Optical and Electronic Information, Huazhong University of Science and Technology, Wuhan, Hubei 430074, China*
*qwzhan@usst.edu.cn

**Abstract:** Spatiotemporal optical vortices (STOVs) carrying transverse orbital angular momentum (OAM) are of rapidly growing interest for the field of optics due to the new degree of freedom that can be exploited. In this paper, we propose cylindrical vector two dimensional STOVs (2D-STOVs) containing two orthogonal transverse OAMs in both x-t and y-t planes for the first time, and investigate the tightly focusing of such fields using the Richards-Wolf vectorial diffraction theory. Highly confined spatiotemporal wavepackets with polarization structure akin to toroidal topology is generated, whose spatiotemporal intensity distributions resemble the shape of Yo-Yo balls. Highly focused radially polarized 2D-STOVs will produce wavepackets towards transverse magnetic toroidal topology, while the focused azimuthally polarized 2D-STOVs will give rise to wavepackets towards transverse electric toroidal topology. The presented method may pave a way to experimentally generate the optical toroidal wavepackets in a controllable way, with potential applications in electron acceleration, photonics, energy, transient light-matter interaction, spectroscopy, quantum information processing, etc.

## 1. Introduction

Photonics orbital angular momentum (OAM) can greatly extend the degree of freedom of light, and have wide applications in high capacity optical communication [1,2], optical tweezers [3-6], super resolution microscopy [7,8], quantum information processing [9,10], optical metrology [11-13], et al. In the past decades, researchers mainly focused on the longitudinal OAM, which is parallel to the propagation direction of the beam [14,15]. However, recent investigations reveal that there also exists transverse OAM, which is orthogonal to the propagation direction of the beam. Theoretically, spatiotemporal optical vortex (STOV) can be created in polychromatic beam by applying spiral phase in the spatiotemporal domain [16]. Subsequently, STOVs are experimentally demonstrated as a result of the nonlinearly arrested self-focusing collapse during extremely high power short pulse filamentation in air [17]. More recently, based on two-dimensional Fourier transformation from the spatial frequency-frequency domain to the spatial-temporal domain, a linear method is developed to experimentally generate STOVs in a controllable way [18,19]. The propagation of STOVs in free space is studied by utilizing a transient grating single-shot super-continuum spectral interferometry to measure the intensity and phase evolutions of STOVs [20].

Due to its intriguing characteristics, increasing attentions have been paid to this emerging area. For example, fast modulation of transverse OAM was demonstrated by embedding two STOVs with different topological charge in one wavepacket, and the temporal separation between the two STOVs is also controllable by adjusting the complex phase patterns applied to the spatial light modulator (SLM) in the pulse shaper [21]. The second harmonic generation of STOV is theoretically and experimentally investigated to reveal the conservation of transverse OAM in such process, namely the spatiotemporal topological charge of the fundamental beam is doubled with the increase of the light frequency [22,23]. Furthermore, the conversion and engineering of transverse OAM in the high harmonic generation of STOV is also studied, discovering the spatially spectral tilt and the fine interference patterns in the produced harmonic spectra, which are caused by the spatial chirp of the fundamental STOV and the spatiotemporal singularity in the high harmonic generation [24]. STOVs with arbitrarily oriented OAM are generated by utilizing the topological defect of the transmission coefficient in a compact device, whose position and dispersion can be tailored through structural symmetry of the device [25]. In addition, the collision of STOVs and spatial vortices are employed to form tilted OAM in a wavepacket [26]. A space to time mapping method is exploited to create STOV by directly modulating the spatiotemporal spiral phase to the chirped pulses via using the relationship between frequency and time [27]. Moreover, the synthesis and propagation of twisted STOVs with controllable partial coherence and transverse OAM are theoretically investigated [28]. Afterwards, STOVs are experimentally generated based on a light source with partial temporal coherence [29]. The physical properties of the generation and propagation of the STOVs are theoretically analyzed based on diffraction theory, revealing the evolution rules of the STOVs during the propagation process in free space [30]. Further, a class of modal solutions are proposed to describe the propagation of STOVs in dispersive media, showing that the OAM is quantized by the STOV symmetry and the group velocity dispersion [31]. The spin-orbit interaction between the transverse spin angular momentum (SAM) and transverse OAM is analyzed in a vectorial Bessel-type STOV based on the local densities and integral values of the SAM and OAM [32].

These findings are expected to inspire new applications that employ the interactions of STOVs with matters. The ability to tightly focus or localize such light-matter interactions is very important for these applications. For this purpose, the subwavelength focusing of scalar STOV is investigated, and a preconditioning method is proposed to tailor the incident spatiotemporal wavepacket to overcome the spatiotemporal astigmatism effect caused by the high numerical aperture (NA) lens, hence the spatiotemporal helical phase distribution can be restored on the focal plane of the lens [33]. The spin-orbit coupling between the longitudinal SAM and transverse OAM in the highly confined circularly polarized STOV results in sophisticated spatiotemporal phase singularity structures, which may render new effects and functions in light-matter interactions [34]. Cylindrical vector (CV) STOVs are a special kind of wavepackets, which simultaneously contain both spatial polarization singularity and spatiotemporal phase singularity, and are experimentally generated recently [35]. The tightly focusing of such kind of STOVs is expected to produce intriguing phenomenon. In this paper, we study the strongly focusing of CV STOVs to reveal the exotic evolution of the polarization distribution in the focused wavepacket on the focal plane of the high NA lens. In order to overcome the collapsing caused by the low degree of symmetry of single transverse OAM to the cylindrical symmetry of the vectorial wavepacket, we mathematically construct CV wavepackets with two orthogonal transverse OAMs, namely two dimensional STOVs (2D-STOVs). And the incident wavepacket will be preconditioned to compensate the spatiotemporal astigmatism effect during the tightly focusing process. Highly confined spatiotemporal wavepackets with polarization distributions akin to toroidal topology can be obtained.

## 2. Cylindrical vector two dimensional STOVs

Without loss of generality, the scalar one dimensional STOV containing transverse OAM with topological charge of +1 in the x-t plane can be described as

$$E(x,y,t) = \frac{\sqrt{x^2+y^2}}{w^2}(x+it)\exp\left(-\frac{x^2+y^2}{w^2}-\frac{t^2}{w_t^2}\right), \qquad (1)$$

where $w$ is the waist radius of the Gaussian profile in spatial domain, and $w_t$ is the pulse half-width at $1/e^2$ of the maximum intensity of the wavepacket in temporal domain. On this basis, the radially polarized incident STOV carrying transverse OAM with topological charge of +1 in both the x-t plane and y-t plane can be expressed as follows

$$\boldsymbol{E}_r^{+1}(x,y,t) = \frac{\sqrt{x^2+y^2}}{w^2}\begin{pmatrix}\cos\phi\cdot\boldsymbol{e}_x\\ \sin\phi\cdot\boldsymbol{e}_y\end{pmatrix}(x+it)(y+it)\exp\left(-\frac{x^2+y^2}{w^2}-\frac{t^2}{w_t^2}\right), \quad (2)$$

where $\phi = \tan^{-1}(y/x)$, $\boldsymbol{e}_x$ and $\boldsymbol{e}_y$ are the respective unit vectors along x- and y- directions. The spatiotemporal distributions of the radially polarized two dimensional STOV are shown in Fig. 1, from which we can see the x-polarized component carries transverse OAM in the y-t plane and the y-polarized component carries transverse OAM in the x-t plane. Supplementary Movie 1 presents the animation of the phase distribution evolutions of the x- and y-polarized components in the corresponding plane. It should be noted that for the x-polarized component, the transverse OAM in the x-t plane is broke due to the sign variation of the factor $\cos\phi$ caused by x coordinate; while for the y-polarized component, the transverse OAM in the y-t plane is broke because of the sign variation of the factor $\sin\phi$ caused by the y coordinate. The entire wavepacket is radially polarized in the x-y plane, forming the polarization singularity. Thus, the radially polarized STOV simultaneously contains one polarization singularity and two spatiotemporal OAM singularities, they are orthogonal to each other.

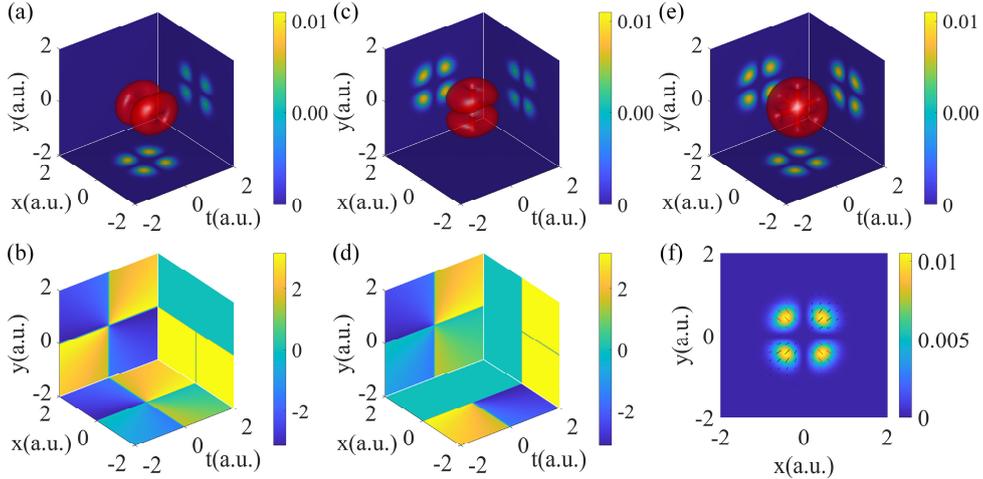

Fig. 1. Spatiotemporal distributions of radially polarized two dimensional STOV. (a) The intensity distributions of the x-polarized component on the three principal coordinate planes (i.e. planes at x=0, y=0, t=0) in the spatiotemporal domain, together with isosurface at 10% of its peak intensity. (b) The phase distributions of the x-polarized component on these three principal planes. (c) The intensity distributions of the y-polarized component on the three principal coordinate planes, together with isosurface at 10% of its peak intensity. (d) The phase distributions of the y-polarized component on these three principal planes. (e) The intensity distributions of the entire wavepacket on the three principal coordinate planes, together with isosurface at 10% of its peak intensity. (f) The polarization distribution of the entire wavepacket in spatial domain. See Movie 1 (Supplementary Information) for animation of phase distribution evolutions of the x-polarized component along the x-axis and that of the y-polarized component along the y-axis.

Similarly, the azimuthally polarized STOV carrying transverse OAM with topological charge of +1 in both the x-t and y-t planes can be given by

$$\boldsymbol{E}_a^{+1}(x,y,t) = \frac{\sqrt{x^2+y^2}}{w^2}\begin{pmatrix}-\sin\phi\cdot\boldsymbol{e}_x\\ \cos\phi\cdot\boldsymbol{e}_y\end{pmatrix}(x+it)(y+it)\exp\left(-\frac{x^2+y^2}{w^2}-\frac{t^2}{w_t^2}\right). \quad (3)$$

Its spatiotemporal distributions are shown in Fig. 2, from which we can see the x-polarized component possesses transverse OAM in the x-t plane, while the y-polarized component carries transverse OAM in the y-t plane. These are opposite to the phenomena in the radially polarized case. The supplementary Movie 2 demonstrates the animation of the phase distribution evolutions of the x- and y-polarized components in the corresponding plane. The entire wavepacket is azimuthally polarized in the x-y plane, and also contains three orthogonal singularities.

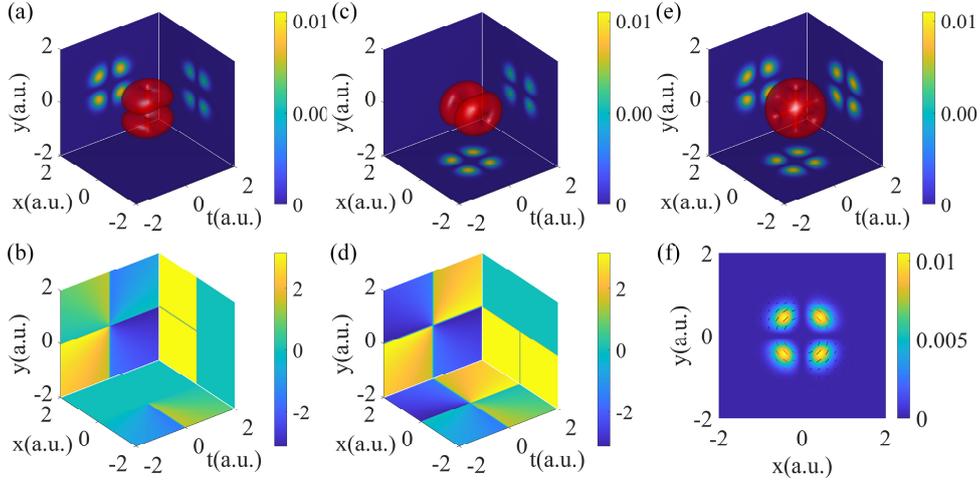

Fig. 2. Spatiotemporal distributions of azimuthally polarized two dimensional STOV. (a) The intensity distributions of the x-polarized component on the three principal coordinate planes, together with isosurface at 10% of its peak intensity. (b) The phase distributions of the x-polarized component on these three principal planes. (c) The intensity distributions of the y-polarized component on the three principal coordinate planes, together with isosurface at 10% of its peak intensity. (d) The phase distributions of the y-polarized component on these three principal planes. (e) The intensity distributions of the entire wavepacket on the three principal coordinate planes, together with isosurface at 10% of its peak intensity. (f) The polarization distribution of the entire wavepacket in spatial domain. See Movie 2 for animation of phase distribution evolution of the x-polarized component along the y-axis and that of the y-polarized component along the x-axis.

## 3. Tightly focusing of CV 2D-STOVs

Previous studies have manifested that the spatiotemporal spiral phase distribution in the STOV will collapse when it is focused by the high NA objective lens due to the spatiotemporal astigmatism effect [33]. In order to prevent the collapse of the focused wavepacket on the focal plane, the incident vectorial spatiotemporal wavepacket can be tailored by employing similar preconditioning method reported in Ref. 33. Hence, the radially polarized incident wavepacket can be preconditioned as

$$\boldsymbol{E}_{rp}^{+1}(x,y,t) = \frac{\sqrt{x^2+y^2}}{w^2}\begin{pmatrix}\cos\phi\cdot\boldsymbol{e}_x\\ \sin\phi\cdot\boldsymbol{e}_y\end{pmatrix}2i(x+t)(y+t)\exp\left(-\frac{x^2+y^2}{w^2}-\frac{t^2}{w_t^2}\right). \quad (4)$$

Similarly, the preconditioned azimuthally polarized incident wavepacket can be written as:

$$\boldsymbol{E}_{ap}^{+1}(x,y,t) = \frac{\sqrt{x^2+y^2}}{w^2}\begin{pmatrix}-\sin\phi\cdot\boldsymbol{e}_x\\ \cos\phi\cdot\boldsymbol{e}_y\end{pmatrix}2i(x+t)(y+t)\exp\left(-\frac{x^2+y^2}{w^2}-\frac{t^2}{w_t^2}\right). \quad (5)$$

The preconditioned radially polarized incident wavepacket is shown in Fig. 3. From Figs. 3(a), 3(c) and 3(e), we can see that all the wavepackets of the x-, y-polarized components, and the entire pulse are split into four detached parts. The intensity of the x-polarized component is mainly in the x-t planes, while the intensity of the y-polarized component is mainly in the y-t planes. According to Figs. 3(b) and 3(d), both of the phase distributions of the x- and y-polarized component are discretized into $0.5\pi$ (yellow areas), $-0.5\pi$ (blue areas) and 0 (green areas). As shown in Fig. 3(f), the preconditioned wavepacket is radially polarized in the spatial domain.

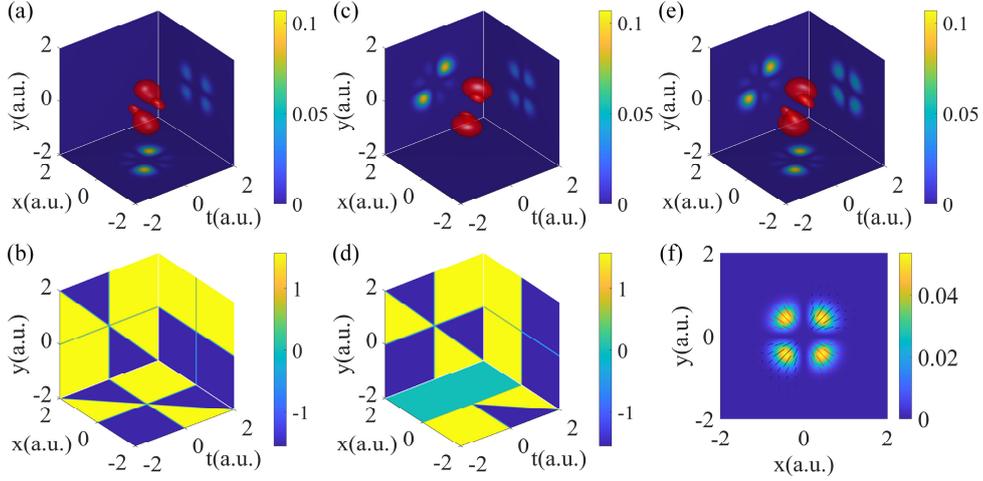

Fig. 3. The preconditioned radially polarized incident wavepacket and its horizontally and vertically polarized components. (a) The intensity distributions of the x-polarized component on the three principal coordinate planes, together with isosurface at 10% of its peak intensity. (b) The phase distributions of the x-polarized component on these three principal planes. (c) The intensity distributions of the y-polarized component on the three principal coordinate planes, together with isosurface at 10% of its peak intensity. (d) The phase distributions of the y-polarized component on these three principal planes. (e) The intensity distributions of the incident wavepacket on the three principal coordinate planes, together with isosurface at 10% of its peak intensity. (f) The polarization distribution of the incident wavepacket in spatial domain.

The preconditioned azimuthally polarized incident wavepacket is shown in Fig. 4. Similar to the former case, the wavepacket of the x-, y-polarized components, and the entire pulse are also split into four detached parts. Different from the radially polarized case, the intensity of the x-polarized component of the azimuthally polarized incident wavepacket is mainly in the y-t planes, while the intensity of its y-polarized component is mainly in the x-t planes. According to Figs. 4(b) and 4(d), both of the phase distributions of the x- and y-polarized component are also discretized into $0.5\pi$ (yellow areas), $-0.5\pi$ (blue areas) and 0 (green areas). Fig. 4(f) clearly shows that the preconditioned wavepacket is azimuthally polarized in the spatial domain.

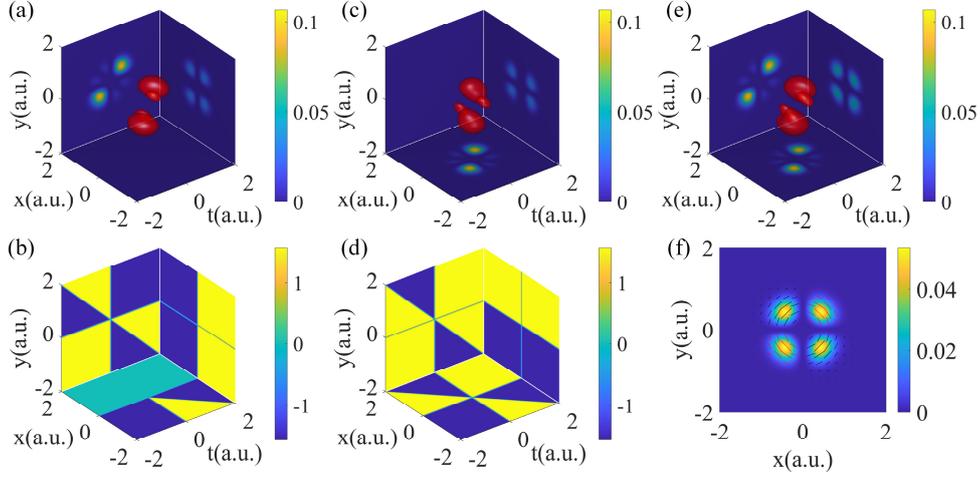

Fig. 4. The preconditioned azimuthally polarized incident wavepacket and its horizontally and vertically polarized components. (a) The intensity distributions of the x-polarized component on the three principal coordinate planes, together with isosurface at 10% of its peak intensity. (b) The phase distributions of the x-polarized component on these three principal planes. (c) The intensity distributions of the y-polarized component on the three principal coordinate planes, together with isosurface at 10% of its peak intensity. (d) The phase distributions of the y-polarized component on these three principal planes. (e) The intensity distributions of the incident wavepacket on the three principal coordinate planes, together with isosurface at 10% of its peak intensity. (f) The polarization distribution of the incident wavepacket in spatial domain.

The schematic diagram to focus the preconditioned incident wavepackets is shown in Fig. 5. In the numerical calculation of focal field, a simplified model is adopted, which ignores the spatiotemporal coupling, so that each temporal slice of the incident wavepacket is assumed to focus on the conjugate temporal position within the focal space. Here, chromatic and other aberrations are also ignored. Subsequently, vectorial Debye diffraction integral formula [36,37] is employed to calculate the focused spatiotemporal wavepacket on the focal plane, which can be expressed as the following:

$$\boldsymbol{E}_f\left(r_f,\Phi,z_f,t\right) = \int_0^\alpha \int_0^{2\pi} \boldsymbol{P}(\theta,\phi) B(\theta) E_\Omega(\theta,\phi,t) \\ \times \exp\{-ik[r_f \sin\theta \cos(\phi-\Phi) + z_f \cos\theta]\} \sin\theta d\theta d\phi \quad , \quad (6)$$

where α is the convergence half angle determined by the NA of the lens. $r_f = \sqrt{x_f^2 + y_f^2}$, $z_f = 0$, $\Phi = \tan^{-1}(y_f/x_f)$, $\boldsymbol{P}(\theta,\phi)$ is the polarization distribution of the refracted field on the spherical surface Ω, $B(\theta)$ is the apodization function of the objective lens. For the sine condition lens, $B(\theta) = \sqrt{\cos\theta}$. $E_\Omega(\theta,\phi,t)$ is the complex amplitude distribution of the refracted field on the spherical surface Ω. Considering the transformation relationship between Cartesian coordinate and spherical coordinate, $E_\Omega(\theta,\phi,t)$ can be written as:

$$E_\Omega(\theta,\phi,t) = 2i\frac{\sin\theta}{w^2}(\sin\theta\cos\phi + t)(\sin\theta\sin\phi + t)\exp\left(-\frac{\sin^2\theta}{w^2} - \frac{t^2}{w_t^2}\right). \quad (7)$$

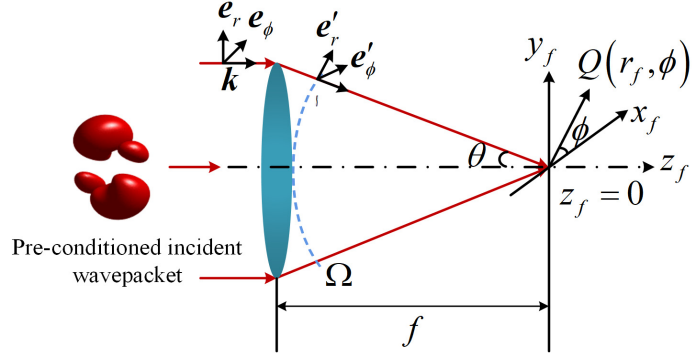

Fig. 5. Schematic diagram of the tightly focusing preconditioned incident wavepacket, $Q(r_f,\phi)$ is an arbitrary observation point on the focal plane of the objective lens. The vector $s$ is the local wave vector of the refracted field in the image space of the lens.

In Fig. 5, $e_r$ and $e_\phi$ are the radial and azimuthal unit vectors in cylindrical coordinate in the pupil plane, respectively. And $e'_r$ and $e'_\phi$ are the radial and azimuthal unit vectors in the spherical coordinate in the image space of the lens. These unit vectors are exploited to illustrate the changes of the polarization direction of radially and azimuthally polarized component of the incident wavepacket after being refracted by the objective lens. As shown in Fig. 5, the radially polarized component of the incident wavepacket are rotated from $e_r$ direction to $e'_r$ direction, which leads to the generation of the z-polarized component in the focused field. The azimuthally polarized component of the incident wavepacket is kept unchanged after being refracted by the lens. These unit vectors can be associated with the three unit vectors ($e'_x$, $e'_y$ and $e'_z$) along the axes of the Cartesian coordinates in the image space of the high NA lens through the following transformation relationships:

$$|e'_r| = |e_r|, \quad |e'_\phi| = |e_\phi|, \tag{8}$$

$$e'_r = \cos\theta\left(\cos\phi\, e'_x + \sin\phi\, e'_y\right) + \sin\theta\, e'_z, \tag{9}$$

$$e'_\phi = -\sin\phi\, e'_x + \cos\phi\, e'_y, \tag{10}$$

Therewith, the polarization distribution $P(\theta,\phi)$ of the refracted beam can be given as [38]

$$P(\theta,\phi) = R^{-1}CRP_0, \tag{11}$$

where $P_0$ is the polarization vector of the incident wavepacket. $R$ accounts for the rotation of the coordinate system around the optical axis, and can be expressed as

$$R = \begin{pmatrix} \cos\phi & \sin\phi & 0 \\ -\sin\phi & \cos\phi & 0 \\ 0 & 0 & 1 \end{pmatrix}, \tag{12}$$

$C$ describes how the polarization is changed on propagation through the lens, which can be written as

$$C = \begin{pmatrix} \cos\theta & 0 & \sin\theta \\ 0 & 1 & 0 \\ -\sin\theta & 0 & \cos\theta \end{pmatrix}. \tag{13}$$

In the case of radially polarized incident wavepacket, $P_0 = (\cos\phi, \sin\phi, 0)^T$, thus $P(\theta,\phi)$ in Eq. (6) can be deduced as:

$$P(\theta,\phi) = \begin{pmatrix} \left[\cos\phi\left(\cos\theta\cos^2\phi+\sin^2\phi\right)+\sin\phi\left(\cos\theta\sin\phi\cos\phi-\sin\phi\cos\phi\right)\right]\cdot\boldsymbol{e}'_x \\ \left[\cos\phi\left(\cos\theta\cos\phi\sin\phi-\sin\phi\cos\phi\right)+\sin\phi\left(\cos\theta\sin^2\phi+\cos^2\phi\right)\right]\cdot\boldsymbol{e}'_y \\ -\sin\theta\left(\cos\phi\cos\phi+\sin\phi\sin\phi\right)\cdot\boldsymbol{e}'_z \end{pmatrix}. \quad (14)$$

$$= \begin{pmatrix} (\cos\theta\cos\phi)\cdot\boldsymbol{e}'_x \\ (\cos\theta\sin\phi)\cdot\boldsymbol{e}'_y \\ (-\sin\theta)\cdot\boldsymbol{e}'_z \end{pmatrix}$$

Similarly, in the case of azimuthally polarized incident wavepacket, $\boldsymbol{P}_0 = (-\sin\phi, \cos\phi, 0)^T$, thus $\boldsymbol{P}(\theta,\phi)$ in Eq. (6) can be derived as:

$$P(\theta,\phi) = \begin{pmatrix} \left[-\sin\phi\left(\cos\theta\cos^2\phi+\sin^2\phi\right)+\cos\phi\left(\cos\theta\sin\phi\cos\phi-\sin\phi\cos\phi\right)\right]\cdot\boldsymbol{e}'_x \\ \left[-\sin\phi\left(\cos\theta\cos\phi\sin\phi-\sin\phi\cos\phi\right)+\cos\phi\left(\cos\theta\sin^2\phi+\cos^2\phi\right)\right]\cdot\boldsymbol{e}'_y \\ -\sin\theta\left(-\sin\phi\cos\phi+\cos\phi\sin\phi\right)\cdot\boldsymbol{e}'_z \end{pmatrix}. \quad (15)$$

$$= \begin{pmatrix} (-\sin\phi)\cdot\boldsymbol{e}'_x \\ (\cos\phi)\cdot\boldsymbol{e}'_y \\ 0 \end{pmatrix}$$

Then, substituting Eq. (14) (or Eq. (15)) and Eq. (7) in to Eq. (6), we can calculate the three dimensional structure of the tightly focused wavepackets corresponding to the radially and azimuthally polarized incident wavepacket, respectively.

## 4. Spatiotemporal structure of highly focused CV 2D-STOVs

In the following numerical simulations of the tightly focused wavepacket, we normalize the spatial size of the incident wavepacket to the NA of the lens, and set NA=0.9, $w = 0.5$ and $w_t = 0.5$.

Firstly, we will present the results of tightly focusing the preconditioned radially polarized incident wavepacket. The intensity and phase distributions of the x-, y- and z-polarized components of the highly confined spatiotemporal field on the focal plane of the lens are shown in Fig. 6. From Figs. 6(a) and 6(c), we can see the wavepacket of the x-polarized component is orthogonal to that of the y-polarized component. As shown in Fig. 6(b), the phase distribution of the x-polarized component in the y-t plane varies anti-clockwise from -π to π, revealing the x-polarized component carries transverse OAM with topological charge of +1 in this plane. Similarly, as shown in Fig. 6(d), the y-polarized component carries transverse OAM with topological charge of +1 in the x-t plane. For the z-polarized component (as shown in Figs. 6(e) and 6(f)), there exist transverse OAMs in both the x-t and y-t planes, causing the holes along the y- axis and x-axis in the wavepacket, respectively. The animation of phase distribution evolution of the z-polarized component in the three orthogonal planes of the spatiotemporal domain are demonstrated in the supplementary Movie 3. From Movie 3, we can see the z-polarized component also carries longitudinal OAM, which leads to the hole along the t-axis in the wavepacket of z-component.

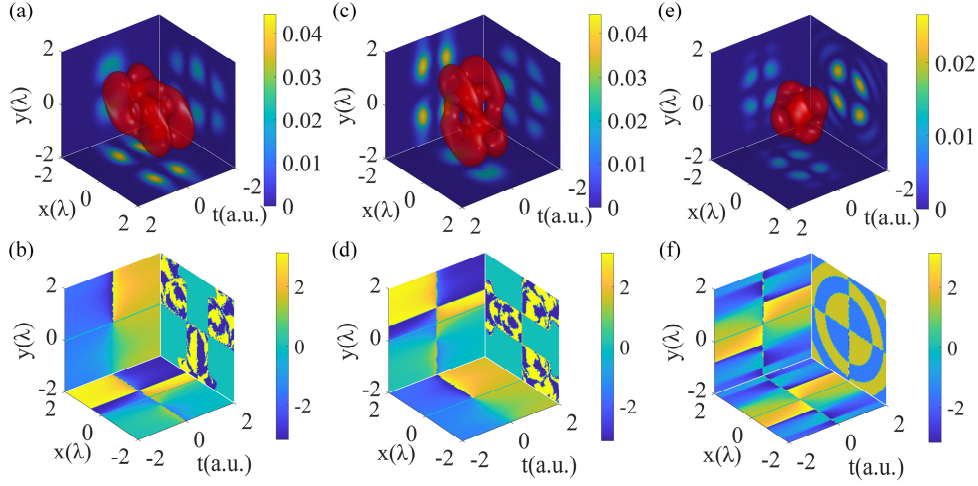

Fig. 6. The intensity and phase distributions of the three orthogonally polarized components of the highly confined spatiotemporal field corresponding to the preconditioned radially polarized incident wavepacket. (a) The intensity distributions of the x-polarized component on the three principal coordinate planes, together with isosurface at 20% of its peak intensity. (b) The phase distributions of the x-polarized component on these three principal planes. (c) The intensity distributions of the y-polarized component on the three principal coordinate planes, together with isosurface at 20% of its peak intensity. (d) The phase distributions of the y-polarized component on these three principal planes. (e) The intensity distributions of the z-polarized component on the three principal coordinate planes, together with isosurface at 20% of its peak intensity. (f) The phase distributions of the z-polarized component on these three principal planes. See Movie 3 for animation of phase distribution evolution of the z-polarized component in the three orthogonal planes.

The intensity distributions of the entire highly confined spatiotemporal field on the three principal coordinate planes together with isosurface at 30% of its peak intensity are shown in Fig. 7(a), from which we can see the focused wavepacket is in doughnut shape. The polarization distributions in the x-y plane at several representative temporal locations are shown in Figs. 7(b)-7(d), the animation of the spatial polarization evolution of the focused wavepacket is given in the supplementary Movie 4. Figure 7(b) presents the intensity distribution superimposed with polarization map near the head of the wavepacket. Figure 7(c) shows the intensity and polarization distributions at the maximum intensity position in the first half of the focused wavepacket. Figure 7(d) gives the intensity and polarization distribution in the center slice of the wavepacket. Since the focused wavepacket is symmetry about the plane at t=0, the slices of the latter half are not depicted. As the slice scans through the focused wavepacket along the t-axis from the head to the end, the polarization distribution in the x-y plane almost maintains as radial polarization. Distinctively, when the slice is near the middle of the focused wavepacket, the spatial polarization distribution degenerates into quadrupole-like polarization, which can be regarded as a special toroidal states [39]; the areas around each polarization singularity are still in radial polarization. Thus, a transverse magnetic (TM) toroidal-like wavepacket with transverse OAM is generated when the preconditioned radially polarized incident wavepacket is tightly focused. The TM toroidal wavepacket has magnetic field tracing the equatorial lines of the torus (perpendicular to the propagation direction) and electric field tracing its meridians (exhibiting electric field components oriented along the direction of wavepacket propagation) [39].

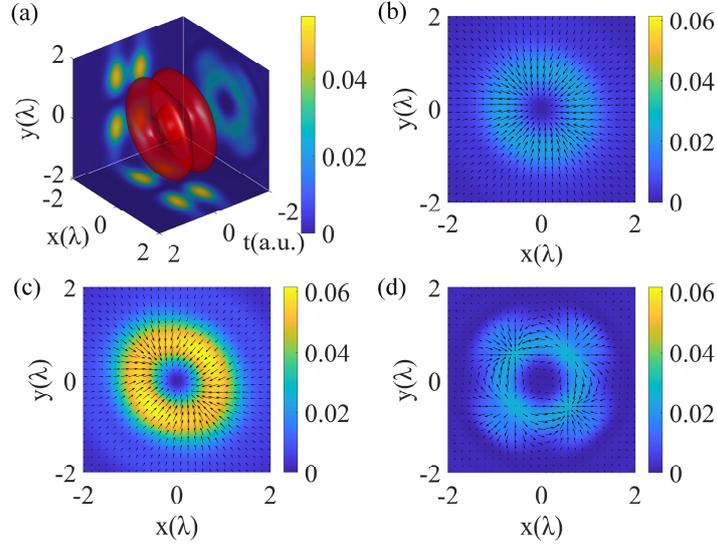

Fig. 7. The intensity and spatial polarization distributions of the whole highly confined spatiotemporal field corresponding to the preconditioned radially polarized incident wavepacket. (a) The intensity distributions on the three principal coordinate planes, together with isosurface at 30% of its peak intensity. The intensity distribution combined with polarization map in the xy planes at (b) t = -0.64, (c) t = -0.4, and (d) t = 0. See supplementary Movie 4 for animation of polarization evolution in the x-y plane as the slice scans along the t-axis.

Then, we discuss the results of tightly focusing the preconditioned azimuthally polarized incident wavepacket. The intensity and phase distributions of the x- and y-polarized components of the highly confined spatiotemporal field on the focal plane of the lens are shown in Fig. 8. Different from the former case, there is no z-polarized component in this tightly focused wavepacket. From Figs. 8(a) and 8(c), we can see the wavepackets of these two polarized components are orthogonal to each other. According to Figs. 8(b) and 8(d), the x-polarized component carries transverse OAM in the x-t plane, while the y-polarized component possesses transverse OAM in the y-t plane.

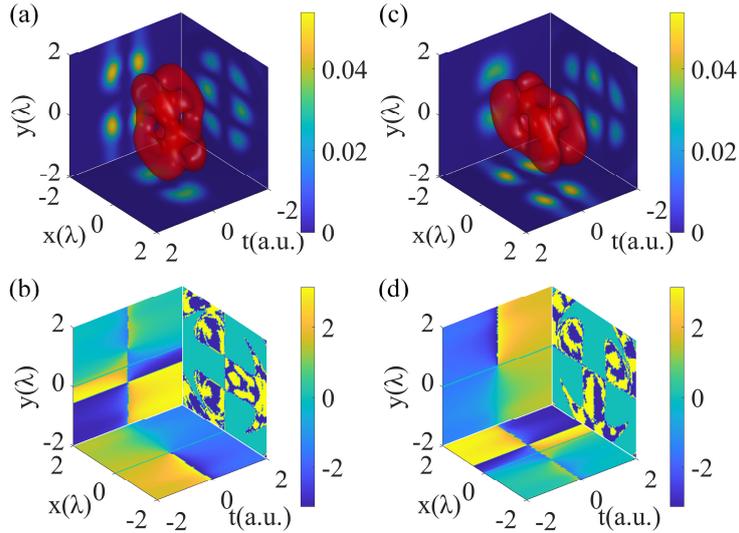

Fig. 8. The intensity and phase distributions of the two orthogonally polarized components of the highly confined spatiotemporal field corresponding to the preconditioned azimuthally polarized incident wavepacket. The intensity distributions of the (a) x-polarized and (c) y-polarized components on the three principal coordinate planes, together with isosurface at 20% of the corresponding peak intensities. The phase distributions of the (b) x-polarized and (d) y-polarized components on these three principal planes.

The intensity distributions of the total highly confined spatiotemporal field on the three principal coordinate planes together with isosurface at 30% of its peak intensity are shown in Fig. 9(a), from which we can see the focused wavepacket is also in doughnut shape. Similar to the former case, the polarization distributions in the x-y plane at several representative temporal locations are shown in Figs. 9(b)-9(d), the animation of the spatial polarization evolution of the focused wavepacket is presented in the supplementary Movie 5. As the slice scans through the focused wavepacket from the head to the end, the spatial polarization distributions almost keep as azimuthal polarization. Specially, when the slice is near the middle of the wavepacket, the polarization distribution also degenerates into quadrupole-like polarization, but still showing azimuthal polarization distribution around each polarization singularity. Therefore, a transverse electric (TE) toroidal-like wavepacket with transverse OAM is generated when the preconditioned azimuthally polarized incident wavepacket is tightly focused. The TE toroidal wavepacket has electric field oscillating perpendicularly to the propagation direction, and its magnetic field possesses longitudinal field components [39].

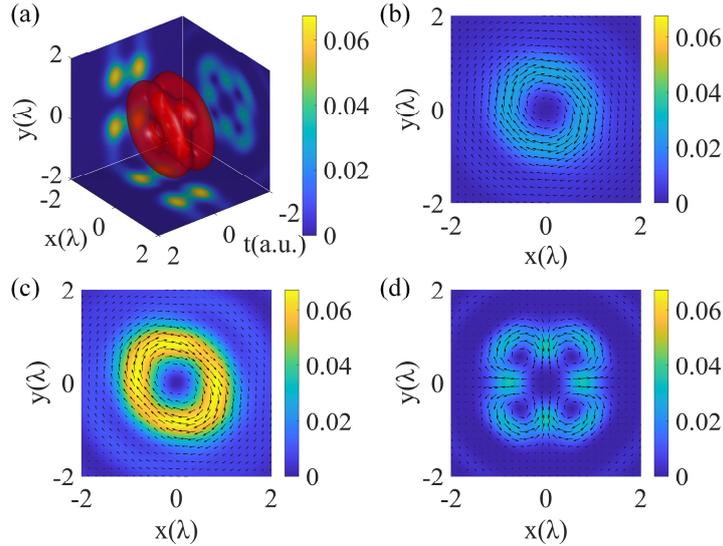

Fig. 9. The intensity and polarization distributions of the whole highly confined spatiotemporal field corresponding to the preconditioned azimuthally polarized incident wavepacket. (a) The intensity distributions on the three principal coordinate planes, together with isosurface at 30% of its peak intensity. The intensity distribution combined with polarization map in the xy planes at (b) t = -0.64, (c) t = -0.48, and (d) t = 0. See supplementary Movie 5 for animation of polarization evolution in the x-y plane as the slice scans along the t-axis

## 5. Conclusions

In conclusion, we study the tightly focusing characteristics of CV 2D-STOV wavepackets and demonstrate unique spatiotemporal structures at the focal plane. Distinctive Yo-Yo ball like spatiotemporal distributions can be realized with cylindrical polarization in the transverse plane. Particularly, as the pulse width approaching the single cycle limit, highly focused radially polarized 2D-STOV wavepacket will form a spatiotemporal structure towards TM toroidal topology while highly focused azimuthally polarized 2D-STOV wavepacket would lead towards TE toroidal topology. The transverse OAMs are carried by each polarization component of the focused wavepacket. However, it should be noted that, in the case of a few cycles of single cycle, the presented vectorial focusing model is over-simplified and many other factors such as the chromatic aberration, lens aberration and spatiotemporal coupling arising from the ultrashort pulse need to be considered. Even so, this work is expected to provide an effective approach to experimentally generate the optical toroidal wavepacket in a controllable way, especially in combination of recent advances in the experimental generation of spatiotemporal optical vortices in free space [18,35]. The unique toroidal-like wavepackets with transverse OAM may find potential applications in light-matter interactions, plasma physics, particle acceleration, energy, and so on.

**Funding**

National Natural Science Foundation of China (92050202, 61805142, 61875245); Shanghai Science and Technology Committee (19060502500); Shanghai Natural Science Foundation (20ZR1437600).

**Disclosures**

The authors declare no conflicts of interest.